\documentclass[11pt]{article}
\usepackage{moriond,epsfig}

\def\Journal#1#2#3#4{{#1} {\bf #2}, #3 (#4)}

\def\PR{{\em Phys. Rept.}}
\def\ARNPS{{\em Ann. Rev. Nucl. Part. Sci.}}

\newcommand{\beq}{\begin{equation}}
\newcommand{\eeq}{\end{equation}}

\def\be{\begin{equation}}
\def\ee{\end{equation}}
\def\bea{\begin{eqnarray}}
\def\eea{\end{eqnarray}}

\def\lsim{\raise0.3ex\hbox{$<$\kern-0.75em\raise-1.1ex\hbox{$\sim$}}}
\begin{document}
\vspace*{4cm}

\title{NUCLEAR STRUCTURE FUNCTIONS AT SMALL $x$\\ FROM INELASTIC SHADOWING AND DIFFRACTION}

\author{ \underline{J.L. ALBACETE}$^{1,2}$, N. ARMESTO$^1$, A. CAPELLA$^3$, A.B. KAIDALOV$^4$  and C.A. SALGADO$^1$}

\address{$^1$Theory Division, CERN, CH-1211 Gen\`eve 23, Switzerland \\
$^2$Dpto. de F\'{\i}sica, M\'odulo C2, Planta Baja, Universidad de C\'ordoba, 14071, Spain\\
$^3$LPT, Universit\'e de Paris XI,
B$\hat{a}$timent 210,F-91405 Orsay Cedex, France\\
$^4$Institute of Theoretical and Experimental Physics, B. Cheremushkinskaya 25, Moscow 117259, Russia }

\maketitle\abstracts{A model for shadowing in nuclear structure functions at small $x$ and small or moderate $Q^2$ is presented using the relation with diffraction that arises from Gribov's reggeon calculus. A reasonable description of the experimental data is obtained with no fitted parameters. A comparison with predictions from other models is  performed.}

The study of nuclear structure functions is a very fashionable subject and has a great importance in the analysis and interpretation of results from heavy ion experiments. At small values of the Bjorken variable $x$ ($\lsim \ 0.01$,
shadowing region), the
structure function $F_2$ per nucleon turns out to be smaller in nuclei than
in a free nucleon [1]. Several explanations
to this shadowing have been proposed. In the rest frame on the nucleus nuclear shadowing can be seen as a consequence of multiple scattering: the incoming virtual photon splits into a colorless $q\bar q$ pair long before reaching the nucleus, and this dipole interacts with typical hadronic cross sections which results in absorption. Multiple scattering can be related to diffraction by means of the AGK rules [2]. Equivalently in a
frame in which the nucleus is moving fast, gluon recombination due
to the overlap of the gluon clouds from different nucleons reduces the
gluon density
in the nucleus [3].

Following the first approach, the $\gamma^*$-nucleus cross
section can be expanded in a multiple scattering series containing
the
contribution from 1, 2,$\dots$ scatterings between the probe and the different nucleons inside nuclei:
\beq
\label{eq1}
\sigma_A = \sigma_A^{(1)}+ \sigma_A^{(2)}+\cdots.
\eeq
$\sigma_A^{(1)}$ is simply
equal to $A\sigma_{{\rm nucleon}}$. The
first correction to the non-additivity of cross sections comes from
the second-order rescattering $\sigma_A^{(2)}$.

To compute it we need
the total contribution which arises from cutting the two-exchange amplitude in
all possible ways (between the amplitudes and the amplitudes themselves in
all possible manners). It can be shown that, for purely imaginary amplitudes, this
total contribution
is identical to minus the contribution from the diffractive cut. Thus
diffractive DIS becomes linked to the first contribution to nuclear shadowing.
The final expression reads
\beq
\label{eq2}
\sigma_A^{(2)}=-4\pi A(A-1)\int d^2b\ T_A^2(b)
\int _{M^2_{min}}^{M^2_{max}}dM^2 \left.
\frac{d\sigma^{\mathcal{D}}_{\gamma^*{\rm p}}}{dM^2dt}\right\vert_{t=0}
F_A^2(t_{min}),
\eeq
with $T_A(b)$ the nuclear profile
function normalized to unity, and
$M^2$ the mass of the diffractively produced system.
 $Q^2$, $x$, $M^2$ and $t$,
or
$x_P=x/\beta$, $\beta=\frac{Q^2}{Q^2+M^2}$ are the usual variables in diffractive DIS. Coherence effects, i.e. the coherence length of the $q\bar q$ fluctuation of
the incoming virtual photon, are taken into account through
\beq
F_A(t_{min})=\int d^2b\ J_0(b\sqrt{-t_{min}})T_A(b),
\label{eq2-1}
\eeq
with $t_{min}=-m_N^2x_P^2$ and $m_N$ the nucleon mass.
This function is equal to 1 at $x\to 0$ and
decreases
with increasing $x$ due to the loss of coherence for $x>x_{crit}\sim(m_NR_A)^{-1}$. The lower integration limit in (\ref{eq2}) (and (\ref{eq6}) below) is $4m_\pi^2\simeq
M^2_{min}=0.08$ GeV$^2$,  while the upper one is taken from the condition:
\beq
\label{eq3}
x_P=x\left(\frac{M^2+Q^2}{Q^2}\right)\leq x_{Pmax}
\Longrightarrow M^2_{max}= Q^2\left(\frac{x_{Pmax}}{x}-1\right),
\eeq
with $x_{Pmax}=0.1$.% this value was used in \cite{model2} motivated by the fact
%that the model there is only valid for $M^2\ll W^2$ or $x_P\ll 1$. In our
%case, variations of $x_{Pmax}$ by a factor 2 do not affect the description
%of nuclear shadowing at $x<0.01$ but the choice $x_{Pmax}=0.1$ is convenient as
%it guarantees the disappearance of nuclear shadowing at $x\sim 0.1$ as experimental data indicate.

Higher order rescatterings are model dependent. Two different ways to unitarize the total cross section have been considered: a Schwimmer unitarization [4] which is obtained from a summation
of fan diagrams with triple Pomeron interactions,
\beq
\label{eq4}
\sigma^{Sch}_{\gamma^*A}=\sigma_{\gamma^*{\rm nucleon}}\int d^2b
\ \frac{AT_A(b)}{1+(A-1)f(x,Q^2)T_A(b)}\ ,
\eeq
and an eikonal unitarization,
\beq
\label{eq5}
\sigma^{eik}_{\gamma^*A}=\sigma_{\gamma^*{\rm nucleon}} \int d^2b
\ \frac{A}{2(A-1)f(x,Q^2)}\left\{1-\exp{\left[-2(A-1)T_A(b)f(x,Q^2)\right]
}\right\},
\eeq
where 
\beq
\label{eq6}
f(x,Q^2)=\frac{4\pi}{\sigma_{\gamma^*{\rm nucleon}}}
\int _{M^2_{min}}^{M^2_{max}}dM^2 \left.\frac{d\sigma^{\mathcal{D}}
_{\gamma^*{\rm p}}}{dM^2dt}\right\vert_{t=0}F_A^2(t_{min})
\eeq
is the key ingredient for shadowing and $\sigma_{\gamma^*{\rm nucleon}}$ and $\left.\frac{d\sigma^{\mathcal{D}}
_{\gamma^*{\rm p}}}{dM^2dt}\right\vert_{t=0}$ have been computed  through $F_2(x,Q^2)$ and $F^{(3)}_{2\mathcal{D}}(Q^2,x_P,\beta)$ for the nucleon taken from CFKS model [5] for $\gamma^*$-p inclusive and diffractive production\footnote{For more details about these derivations, see [6].}.
Thus, the region of applicability of our model is that of CFKS model
i.e. that of small $x\lsim 0.01$ and small or moderate
$Q^2 \lsim 10$ GeV$^2$, including photoproduction. Let us notice
that our model is devoted to the small $x$ region and therefore no antishadowing
or any other effects relevant for $x\ge 0.1$ have been introduced. As at low $x$ the contribution of valence quarks is negligible no distinction is done between protons and neutrons. Also the
experimental results have been isospin-corrected, so the comparison of the
results of the model with those of experiment is legitimate. 

Shadowing in nuclei is usually studied by ratios of cross sections per
nucleon for
different nuclei, defined as
\beq
\label{eq7}
R(A/B)=\frac{B}{A}\frac{\sigma_{\gamma^{*}A}}{\sigma_{\gamma^{*}B}}.
\eeq

In the simplest case of the ratio over nucleon (equivalent to proton at small
$x$ where the valence contribution can be neglected), we get:
\beq
\label{eq8}
R^{Sch}(A/{\rm nucleon})=\int d^2b\ \frac{T_A(b)}{1+(A-1)f(x,Q^2)T_A(b)}\ ,
\eeq
\beq
\label{eq9}
R^{eik}(A/{\rm nucleon})=\int d^2b
\ \frac{1}{2(A-1)f(x,Q^2)}\left\{1-\exp{\left[-2(A-1)T_A(b)f(x,Q^2)
\right]}\right\}.
\eeq
From Eqs. (8) and (9) it is evident that our model also predicts the evolution of shadowing with centrality.

Comparison of our predictions with experimental data at small $x$
from E665 [7,8] is shown in Fig. 1. 
The agreement with data
is quite reasonable
taking into account that no parameters have been fitted to reproduce the data. Comparison with other set of data has also been performed (see [6]) obtaining a nice agreement as well.

\begin{figure}[ht]
\begin{center}
\psfig{figure=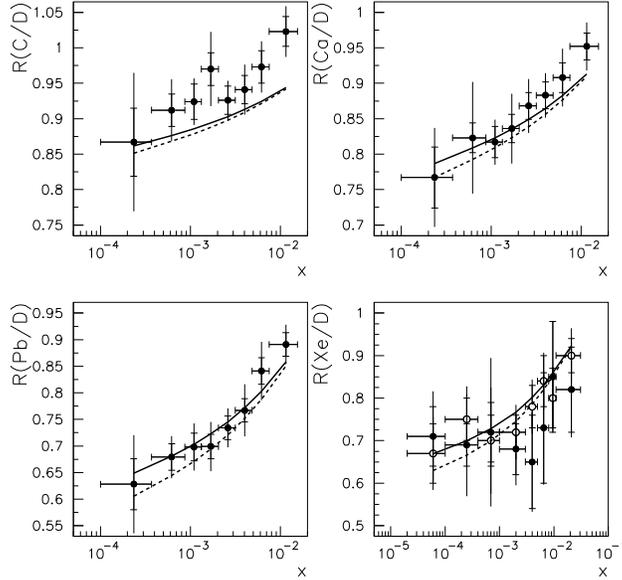,height=3.5in}
\end{center}
\caption{Results of the model using Schwimmer (solid lines) and
eikonal (dashed lines) unitarization compared with experimental data, for the ratios C/D, Ca/D, Pb/D [7] and Xe/D [8].} 
\label{fig1}
\end{figure}

In Fig. 2 a comparison of the results of our model with those of other models is shown, for $Q^2=3$ GeV$^2$. It can be seen that the results of different models agree within $15\%$ at $x=0.01$ where experimental data exist, while they differ up to a factor $0.6$ at $x=10^{-5}$. Future measurements of $F_{2A}(x,Q^{2})$ in lepton-ion colliders with a $10 \%$ of sensitivity will discriminate between different models.

\begin{figure}[ht]
\begin{center}
\psfig{figure=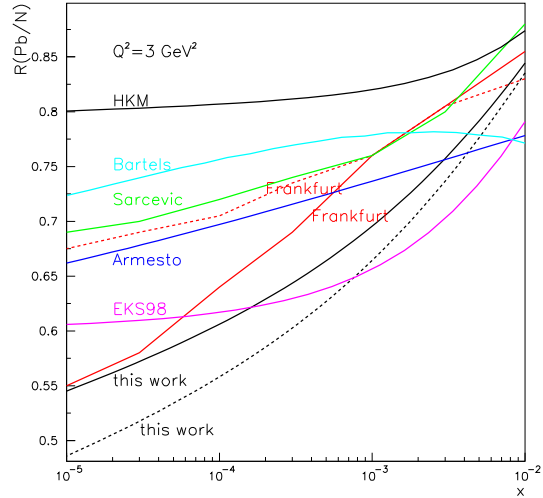,height=3.0in}
\end{center}
\caption{  Comparison of the results of our model
using Schwimmer (solid lines) and
eikonal
(dashed lines) unitarization for the ratio Pb/nucleon
with other models, versus $x$ at fixed $Q^2=3$ GeV$^2$. HKM are the results from [9], Sarcevic from [10], Bartels from [11], Frankfurt from [12],
Armesto from [13] and EKS98 from [14].}
\label{fig2}
\end{figure}

\section*{Conclusions}
A simple model for nuclear shadowing based in the relation between multiple scattering and diffraction provided by AGK rules has been presented. In this way the study of Low $x$ Physics at HERA gets linked with
that of nuclear structure functions at future lepton-ion colliders and
with Heavy Ion Physics at RHIC and LHC. 

\

\end{document}